\documentclass[12pt,review,number]{elsarticle} 
\pdfoutput=1
\usepackage{graphicx}
\usepackage{natbib}

\journal{Computational Materials Science}
\begin{document}             % End of preamble and beginning of text.

\begin{frontmatter} 

  \title{Improved Calculation of Vibrational Mode Lifetimes
      in Anharmonic Solids - Part I: Theory}

 \author{Doyl Dickel}
 \author{Murray S. Daw\corref{cor1}}
 \ead{daw@clemson.edu} 
 \cortext[cor1]{Corresponding author}
 \address{Dept of Physics \& Astronomy, Clemson University, Clemson,
  SC 29634}

  \date{\today} 
 
  \begin{abstract}
    We propose here a formal foundation for practical calculations of
    vibrational mode lifetimes in solids. The approach is based on a
    recursion method analysis of the Liouvillian. From this we derive
    the lifetime of a vibrational mode in terms of moments of the
    power spectrum of the Liouvillian as projected onto the relevant
    subspace of phase space. In practical terms, the moments are
    evaluated as ensemble averages of well-defined operators, meaning
    that the entire calculation is to be done with Monte Carlo. These
    insights should lead to significantly shorter calculations
    compared to current methods. A companion piece presents numerical
    results.
  \end{abstract} 
 

\end{frontmatter}

\section{Introduction}

The state-of-the-art in calculating vibrational mode lifetimes in
solids is surprisingly under-developed. While well-tested, reliable,
and efficient means allow a determination of the electronic structure
of newly discovered and hypothetical crystals --- as least at the
level of mean-field such as LDA --- a comparable technique does not
exist for determining mode lifetimes or its related property of the
lattice thermal conductivity. The aim of this work is then to provide
a fundamental approach that can be used to calculate in
a practical way the intrinsic mode lifetimes of solids. \\

Two methods are used currently for calculating the intrinsic mode
lifetimes: the first combines first-principles calculations of force
constants with standard rate theory, the second utilizes molecular
dynamics (MD) simulations based on inter-atomic potentials. \\ 

The very recent work of Broido \emph{et al.}\cite{broido07}
combines Boltzmann rate formalism with LDA calculations of harmonic
and 3rd-order anharmonic force constants. Without fitting parameters,
they obtain excellent agreement with measurements of intrinsic lattice
thermal conductivity in Si and Ge from 100-300$^\circ$ K. While mode
lifetimes were not directly computed, they could have been by the same
technique, with some additional expense. Quantitative estimates of the
fourth-order rates showed them to be negligible compared to
third-order. The agreement is indeed encouraging. However, the
calculations are quite demanding, and the demands increase
significantly with more complex unit cells. Time and memory
bottlenecks preclude consideration of bulk materials with
larger unit cells. \\

The second technique, based on molecular dynamics (MD) begins with the
Green-Kubo relation\cite{kubo91}, which expresses the lifetime
$\tau_k$ in terms of the mode auto-correlation function $\chi_k(t)$
\begin{equation}
\tau_k = \int_{-\infty}^{+\infty} dt\ \chi_k(t)
\label{eq:GKform}
\end{equation}
where 
\begin{equation}
\chi_k (t) = \frac{\langle \delta n_k(0)\ \delta n_k(t)
  \rangle}{\langle \delta n_k(0)^2 \rangle} 
\label{eq:chik}
\end{equation}
\noindent where $k$ is the mode index (wave-vector), $\delta n$ is
the fluctuation of the occupancy factor, and the angular brackets
indicate averages over the equilibrium distribution in
phase-space.\footnote{The time integral is long compared to the
  times involving energy transfer but is shorter than the Poincare
  recurrence time.} First-principles MD simulations are possible, but
the MD times must be long compared to the lifetimes, which is often
impractical, especially because cell-size effects require the use of
rather large simulation cells. Instead, the G-K relation is usually
restricted to systems where there exists a reliable inter-atomic
potential\cite{dong01,bodapati06b} such as Si and C. That constraint
pretty much limits the studies to those materials which are already
well-studied. Very few new materials are
amenable to this approach. \\

More than twenty years ago, mode lifetimes were calculated by Ladd
\emph{et al.}\cite{ladd86} using MD for Lennard-Jones potentials for
moderate-sized cells (864 atoms). This straightforward calculation
gave surprising results for the lifetimes vs. $k$ and temperature,
which were not explained. To the present authors, this work seems
fundamental and needs to be pursued further. With advances in
computers it should be possible to do significantly larger cells
today. It is also a surprise that --- as far as we can determine ---
no further work has been done
along these lines. \\

The analysis of vibrational mode lifetimes was encompassed as part of
the general theory of energy dissipation in dynamical systems begun in
the seminal work of Van Hove\cite{hove55,hove59}, and further
developed by Prigogine and
co-workers\cite{brout56,prigogine59a,prigogine59b,resibois61,prigogine61,resibois63,prigogine68},
and along similar lines by Zwanzig\cite{zwanzig60,zwanzig61,zwanzig66}
and Mori\cite{mori65}. The basic formalism considered the evolution of
classical dynamical systems using the
Liouvillian\cite{koopman31,koopman32}. Zwanzig and Mori's
``projector-operator'' approach was applied by Wilson and Kim
specifically to the problem of mode lifetimes and lattice thermal
conductivity\cite{wilson73}. While the formalism provided insights
into the mechanism of equilibration of energy, there have been no
quantitative predictions of mode lifetimes using this formalism (at
least as far as we can find). \\

A recent breakthrough has been made in extracting the long-term
dynamics within the Liouvillian formalism by applying the recursion
method\cite{haydock95,haydock99,haydock06}. Haydock, Nex, and
Simons\cite{haydock99} have recently proposed a practical scheme for
calculating macroscopic rates from the resolvent of the Liouvillian.
In this view, dissipation results from the flow of energy from large
to small scale structures in phase space. By using the recursion
method, along with careful considerations of the required analytical
properties of the resolvent, Haydock and company investigated a way of
extracting the long-time behavior from a finite amount of information
about the resolvent. This new insight forms the inspiration and basis
of the current work. \\ 

In this paper, we briefly review the formalism which relates the
Liouvillian to mode lifetimes, and then demonstrate the application of
the recursion method to this problem. On that basis, the problem
becomes one of analyzing low moments of the power spectrum of the
Liouvillian to extract the correct long-time dynamics. We show
some results of the formalism. \\

These ideas are somewhat similar to the work of Tankeshwar \emph{et
  al.}\cite{tankeshwar89a, tankeshwar89b}, who examined energy flow
in a Lennard-Jones liquid using a continued fraction representation of
the memory function. Unlike that work, in the present case we make
explicit use of the Liouvillian formalism, so that the constants in
the continued fraction are related directly to the moments of the
Liouvillian. Also, we can take advantage of the recent work on the
recursion relation in obtaining guidance for terminating the
recursion. These two additions make it possible for us to flesh out a
more complete formalism. And then, there is the obvious difference
that we are dealing with solids here. \\

A companion piece\cite{daw09b} to this paper will lay out the results of applying
the formalism to a simple model of vibrations in solids, in both 1 and
3 dimensions. In the context of the numerical simulations, we will be
able to address definitively the question of convergence which comes
up in the recursion method. \\

\section{Mode Lifetimes and the Green-Kubo Formula}

In the Green-Kubo (GK) approach, the lifetime of an individual
vibrational mode (labelled by wave-vector $k$) is given by
Eq.~\ref{eq:GKform}, where the mode auto-correlation function
$\chi_k(t)$ is given by Eq.~\ref{eq:chik}. In $\chi_k$, $\delta n$ is
the fluctuation of the occupancy factor

\[  \delta n_k = n_k - \langle n_k \rangle \] 

\noindent and where the angular brackets indicate averages over the equilibrium
distribution in phase-space:

\begin{eqnarray}
\langle A \rangle & = & Z^{-1} \int d\Gamma\ e^{-\beta H(\{p_i\},\{q_i\})}\ A(\{p_i\},\{q_i\})  \\
Z  & = & \int d\Gamma\ e^{-\beta H(\{p_i\},\{q_i\})}
\label{eq:ensembleavg}
\end{eqnarray}
with phase space differential volume 
\[ d\Gamma \equiv \prod_i dp_i dq_i \] 

In the auto-correlation function, the appearance of time ($t$) in the
ensemble average can be given the following realization for the
purposes of computation: a point in phase space is selected (for
example, using Monte Carlo or ``MC'') from the equilibrium
distribution, and this point is then taken as the initial condition
for dynamical evolution (using molecular dynamics or ``MD'') for a
time $t$. The product of the fluctuations at these two times is then
averaged over the distribution. The resulting auto-correlation should
decay with time as a result of anharmonicities in the interactions.
The denominator of the auto-correlation is fixed so that the function
goes to 1 as $t \rightarrow 0$. Then the area under the
auto-correlation curve is the desired mode lifetime. In this way, the
GK approach clearly mixes equilibrium and dynamical features. \\

The exact form of the time dependence of the auto-correlation is
subject in some way to details of the interactions but some properties
are general. For example, time-reversal symmetry means that the
auto-correlation is symmetric in $t$, and also that
\begin{equation}
\dot{\chi}(0) = 0
\label{eq:chidot0}
\end{equation}
One might expect on general grounds\cite{prigogine68} for $\chi(t)$
to decay exponentially at very long times. However, the behavior
observed at intermediate times in most applications is more
complicated, and the lifetime in Eq.~\ref{eq:GKform} is more strongly
dependent on $\chi(t)$ at intermediate times. \\

A property of the equilibrium averages which will be helpful is to
know that the equilibrium averages are themselves independent of $t$:
\[ \langle A(t) \rangle = \langle A(0) \rangle \]
from which it follows immediately that $ \langle \dot{A}(t) \rangle =
0$ for any function $A$ of phase space. As a special case we see that 
\begin{equation}
\dot{\chi}(0) = \langle \delta \dot{n}(0)\  \delta n(0) \rangle = 0
\label{eq:nndot}
\end{equation}
consistent with our previous observation. \\

\section{Canonical Transformations}

The expression of $\delta n_k$ for vibrating solids in terms of
particle coordinates and momenta ($q_i$,$p_i$) is accomplished via a
series of two canonical coordinate transformations: a transformation
to normal modes is followed by the Hamilton-Jacobi
transformation to action-angle variables. \\

It is revealing to recall the original arena in which the action-angle
transformation and Hamilton-Jacobi\cite{goldstein80} theory were
developed, which was to study perturbations of planetary orbits by
other planets. In that case, one has possibly strong effects of other
planets, but there remains an underlying periodic nature to the
motion. In the action-angle formalism, the various planets execute
nearly periodic orbits while altering the action and relative phases
of the other orbits. This coupled periodic orbital motion is clearly
quite similar to the case of coupled anharmonic vibrations, where the
underlying periodic motion of the normal modes is influenced by
possibly strong coupling to other normal modes. So even in the
presence of strong coupling, the action-angle transformation can be
very helpful. Perhaps this was the insight that lead to Prigogine's
interest\cite{prigogine68}. At any rate, his work clearly shows the
utility of the transformation. \\ 

The simple SHO hamiltonian involving coordinate and momentum variables
($q$,$p$) 
\[ H(p,q) = \frac{p^2}{2 m} + \frac{k q^2}{2} \] 
can be transformed to action-angle variables ($S$,$\alpha$) by a
transformation involving an arbitrary frequency $\omega^X$
\begin{eqnarray}
q & = & \sqrt{ 2 S/( m \omega^X)} \sin{\alpha} \\
p & = & \sqrt{ 2 S m \omega^X} \cos{\alpha} 
\label{eq:HJ}
\end{eqnarray}
to
\[ K(S,\alpha) = S ( \omega^X \cos^2{\alpha} + \frac{k}{m \omega^X}
\sin^2{\alpha} ) \]
with the resulting equations of motion:
\begin{eqnarray}
\dot{\alpha} & = & \omega^X \cos^2{\alpha} + \frac{k}{m \omega^X}
\sin^2{\alpha} \\
\dot{S} & = & -2 S ( -\omega^X + \frac{k}{m \omega^X}) \sin{\alpha}
\cos{\alpha} 
\end{eqnarray}
The transformation is canonical for any value of $\omega^X$, but the
equations of motion simplify greatly if the transformation frequency
matches the harmonic frequency ($\omega^X = \omega^H =
\sqrt{\frac{k}{m}}$). In that case, $S$ is a constant of the motion
and $K = \omega S$, from which one sees the natural relationship $n =
S/\hbar$.\footnote{Planck's constant appears unnaturally here because
  of the interpretation of $n$ as a quantum number. We deal in this
  proposal only with behavior in the classical realm, but the use of
  occupation number in favor of action is overwhelmingly prevalent.
  Formally more consistent would be to define $\chi_k(t)$
  (Eq.~\ref{eq:chik}) in terms of $S_k$ rather than $n_k$, but
  $\chi_k$ is unaffected anyway.}
Given that condition on $\omega^X$, the angle $\alpha$ evolves
uniformly in time ($\alpha =
\omega^H t$). \\

It is instructive to note that if the transformation frequency does
\emph{not} match the harmonic frequency, the resulting dynamical
variables are still periodic, and their averages over a period of the
motion are $\overline{\dot{S}} = 0$ and $\overline{\dot{\alpha}}
\approx \omega^H$ if $\omega^X$ is in the neighborhood of $\omega^H$.
\\

The action-angle transformation is canonical regardless of the
hamiltonian. It could also be applied to, for example, the anharmonic
hamiltonian 
\[ H(p,q) = \frac{p^2}{2 m} + \frac{k q^2}{2} + \lambda q^4 \] The
motion of the this anharmonic system is yet periodic~\footnote{As long
  as $\lambda \geq 0$.}, so that the angle $\alpha$ will have an
overall linear behavior in time with a periodic oscillation
superimposed. Now the transformation frequency can be chosen to
simplify the equations, but $\alpha$ cannot be made simply linear in
time. Instead, the overall slope of $\alpha$ has the meaning of a
quasi-harmonic frequency for the anharmonic motion, so that, if we
average over a period of the motion, $\overline{\dot{\alpha}} =
\omega^Q$. This observation will come in handy when we tackle the
vibrating anharmonic lattice problem, where we will use a similar
insight to find the quasi-harmonic frequency of a mode interacting
with other modes in an ensemble. Even in this case, as we shall see,
the transformation frequency $\omega^X$ can be chosen to obtain some
convenience. \\

The action-angle formalism can be applied easily to the normal modes
of an anharmonic vibrating network. Consider, for example, a $1D$
anharmonic chain whose hamiltonian is
\begin{equation}
H = \frac{1}{2} \sum_n p_n^2 + \sum_{n} V(u_{n,n+1})
\label{eq:H1D}
\end{equation}
where 
\begin{equation}
V(d) = d^2/2 + d^4/24 
\label{eq:V1D}
\end{equation}
The transformation to normal mode variables $q$ and $\pi$ 
\begin{eqnarray}
q_k & = & \frac{1}{N} \sum_n u_n e^{-i k n} \\
\pi_k & = & \frac{1}{N} \sum_n p_n e^{-i k n} 
\end{eqnarray}
would decouple the modes in the harmonic case (drop the quartic term in
$V$). \\

Even in the anharmonic case, it is useful now to go from normal modes
to action-angle variables, as was done previously. 
\begin{eqnarray}
S_k & = & \frac{N}{2 \omega_k^X} | \pi_k + i \omega_k^X q_k | ^2  =
\hbar n_k \\
\alpha_k & = & \arg{(\pi_k + i \omega_k^X q_k)}
\end{eqnarray}
In the present case, the anharmonicity of $V$ prevents $S_k$ from
being a constant of the motion for any $\omega_k^X$; which is to say
that the anharmonicity is responsible for fluctuations in the
occupation of modes. We also find that the motion is quasi-periodic,
with $\alpha$ having an overall linear progression with superimposed
oscillations. Now it makes sense to define the averages of quantities
in terms of the ensemble, so that we can find the quasi-harmonic
frequency (which depends on both temperature and wave-vector) from:
\[ \langle \dot{\alpha}_k \rangle = \omega_k^Q(\beta) \] Generally it
is expected that at low temperatures, the quasi-harmonic frequency
will become the harmonic frequency of the lattice and deviate as the
temperature increases.  \\

\begin{figure}[h] % figure placement: here, top, bottom, or page
   \includegraphics[width=6in]{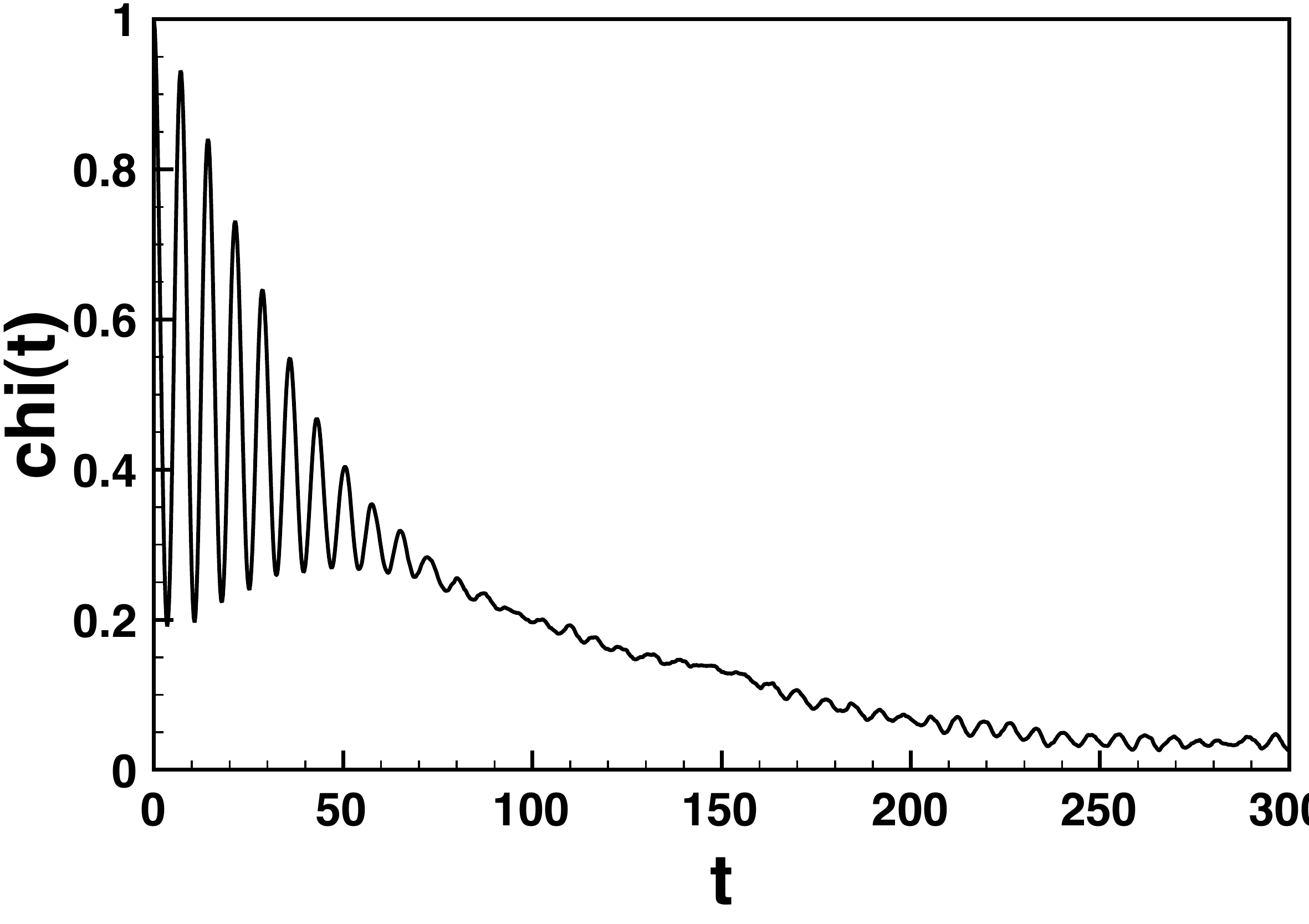} 
   \caption{\emph{A typical occupation auto-correlation (see
       Eq.~\ref{eq:chik}). This example was calculated for a simple
       anharmonic vibrating lattice model (discussed in more detail in
       Ref.~\cite{daw09b}). The frequency of the transform $\omega^X$
       (see Eq.~\ref{eq:HJ}) was fixed to match the harmonic frequency
       of a typical normal mode of the lattice at low
       temperatures. The temperature was then raised substantially for
       this calculation to illustrate the effect of not adjusting the
       transformation frequency to account for temperature shift of
       the mode frequency.}}
  \label{fig:ChiAtOmega0}
\end{figure}

At this point, the transformation frequency $\omega_k^X$ is yet
arbitrary. We will find in the following (departing from Prigogine's
approach) that the most convenient value of the transformation
frequency will be $\omega_k^X = \omega_k^Q(\beta)$ (an implicit
equation for $\omega^X$). This choice is made by considering the
structure of the auto-correlation function and the resulting lifetime
(area under the auto-correlation function). We show in
Fig.~\ref{fig:ChiAtOmega0}
a plot of the auto-correlation function at elevated temperature. (The
data are taken from calculations on a $1D$ anharmonic chain, but the
behavior is generic). In this figure, the transformation frequency was
arbitrarily chosen to be the harmonic (that is, low-temperature)
frequency of the mode and kept at this value even for the analysis of
the dynamics at higher temperature (note that the transformation does
not affect the real dynamics --- as expressed in the original
coordinates --- but only the interpretation in terms of the
transformed coordinates). In this instance, we see that the
auto-correlation function decays as expected but also has oscillations
that complicate the analysis. \\ 

\begin{figure}[h] %  figure placement: here, top, bottom, or page
   \includegraphics[width=6in]{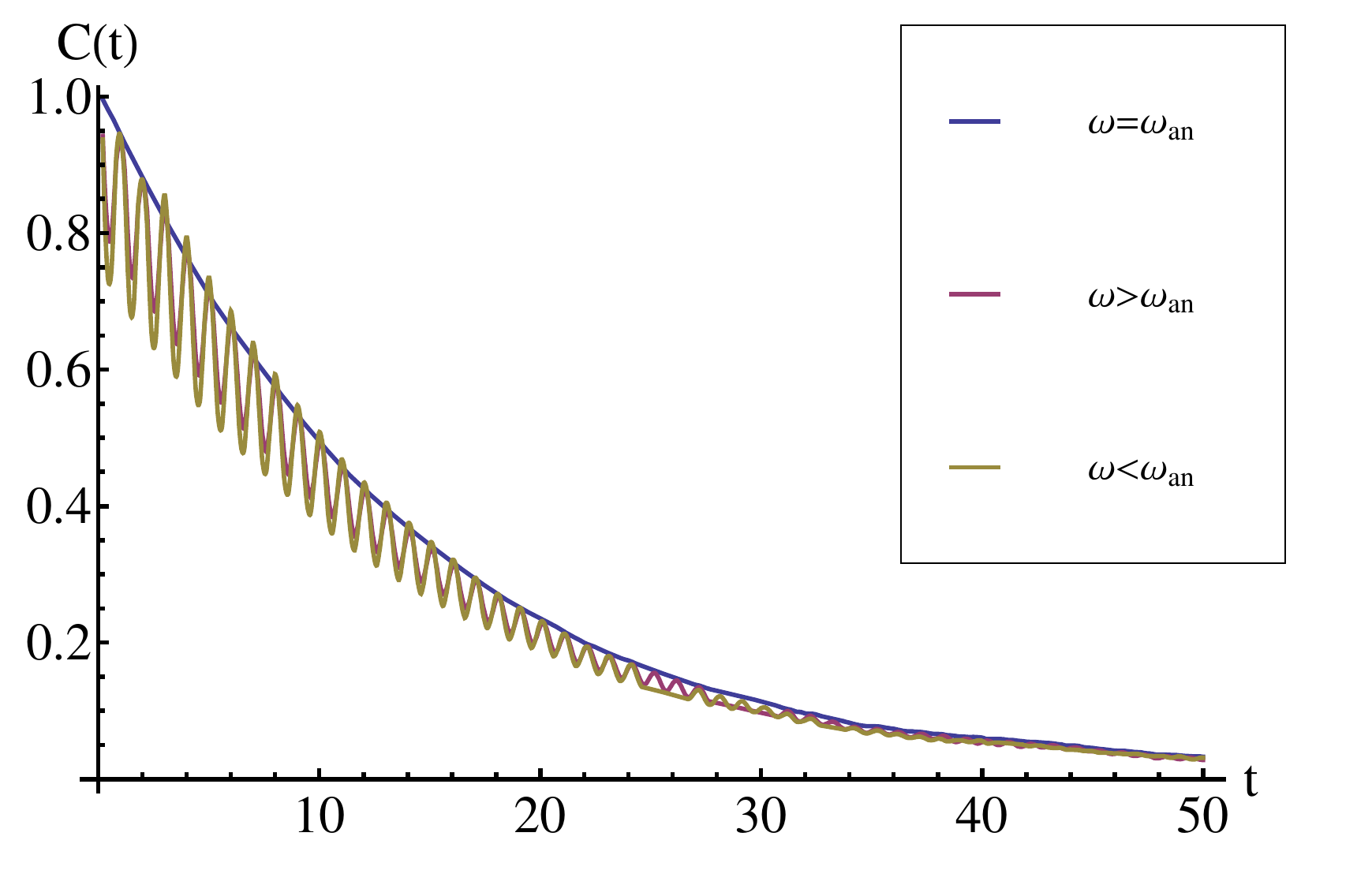} 
   \caption{\emph{The occupation auto-correlation at elevated
       temperature for three values of the transformation frequency
       $\omega^X$. For the top curve, the frequency has been matched
       to the temperature-dependent quasi-harmonic frequency of the
       mode. For the other two curves, the frequency is slightly above
       or below that of the quasi-harmonic frequency. With the
       frequency properly matched to the quasi-harmonic frequency, the
       auto-correlation is smooth. Going off-frequency introduces
       oscillations into the auto-correlation, with the on-frequency
       curve forming a maximum envelope for the oscillations. Clearly
       the area under the curve will be maximum for the
       frequency-matched case. Numerical results are displayed for a
       particular, simple model of anharmonic vibrating
       lattice\cite{daw09b}, but illustrate a general behavior.}}
 \label{fig:ChiVsOmega}
\end{figure}

We can see how the choice of transformation frequency affects the
auto-correlation function in Fig.~\ref{fig:ChiVsOmega},
which shows the function for three values of $\omega^X$. We see that
there exists one value, interpreted as the quasi-harmonic frequency
for the mode, for which the oscillations in the auto-correlation
function vanish, leaving a smoothly decaying function. Choosing the
transformation frequency to be on either side of the quasi-harmonic
frequency introduces oscillations into the auto-correlation. From
this, we see that oscillations in the mode auto-correlation are
artifacts of how we define the transform. If we set the transformation
frequency to be the natural frequency of the system (which is the
temperature-dependence quasi-harmonic frequency), then
the auto-correlation is quite simple. \\

\begin{figure}[h] %  figure placement: here, top, bottom, or page
   \includegraphics[width=6in]{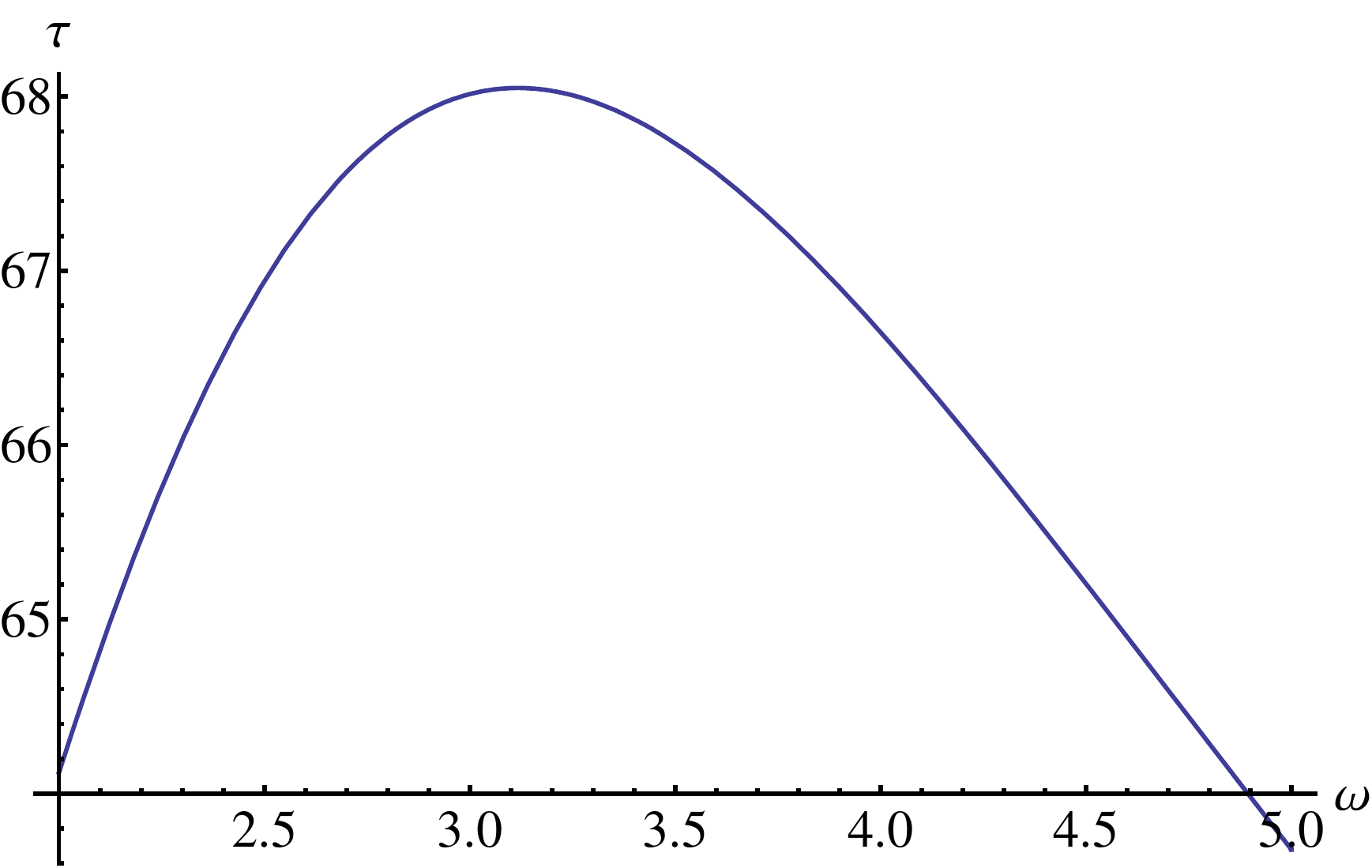} 
   \caption{\emph{Mode lifetime as a function of the transformation
       frequency $\omega^X$ (see Eq.~\ref{eq:HJ}). As demonstrated in
       Fig.~\ref{fig:ChiVsOmega}, the area under the curve (which is
       identified as the mode lifetime) is maximized by matching the
       transformation frequency to the temperature-dependent,
       quasi-harmonic frequency of the mode. Numerical results are
       displayed for a particular, simple model of anharmonic
       vibrating lattice, but should be general.}}
 \label{fig:TauVsOmega}
\end{figure}

An alternative view gives a similar conclusion. Consider that the mode
lifetime ($\tau$) depends formally on the choice of $\omega^X$ used to
define the action-angle transformation. In Fig.~\ref{fig:TauVsOmega}, 
we see that $\tau$ is a concave function of the transformation
frequency. This is quite easily understood in relation to the previous
observations about the auto-correlation function, noting that $\tau$
is the area under the function. We note in Fig.~\ref{fig:ChiVsOmega}
that the oscillations are always below the decaying envelope, so that
zeroing out the oscillations will maximize the area. Thus, choosing
the transformation frequency to match the quasi-harmonic frequency of
the mode also maximizes the calculated lifetime. In the vicinity of
the quasi-harmonic frequency, the lifetime is relatively insensitive
to the transformation frequency. We see then that this definition of
the transformation gives a natural definition for the (quasi-harmonic)
frequency and lifetime. Fixing the transformation frequency in this
manner, we see that \emph{the occupation
  auto-correlation function is a monotonic, decaying function.} This
property will greatly simplify the following analysis. \\

\section{The Liouvillian and Mode Lifetimes}

The evolution of dynamical variables, such as the occupation $\delta
n_k$ as defined in the previous section, are governed by Hamilton's
equations, which are, for anharmonic systems, naturally non-linear and
correspondingly difficult to treat. However, the lifetime of the mode
in a solid in equilibrium is expressed as an ensemble property, as
defined in the auto-correlation Eq.~\ref{eq:chik}. The lifetime is
therefore an ensemble property, and reflects the dynamical behavior of
different parts of the distribution function. The dynamics of
distributions in phase space, even those systems with anharmonicity,
is governed by the Liouvillian, which is a linear operator. That is,
any function $f$ on phase space evolves according to
\[ \frac{\partial f(p,q,t)}{\partial t} = -i \hat{L} f \]
where the (hermitian) Liouvillian operator
\[ \hat{L} = i\{H, \} = 
i \sum_l (
\frac{\partial H}{\partial q_l} \frac{\partial}{\partial p_l} - 
\frac{\partial H}{\partial p_l} \frac{\partial}{\partial q_l} )
\]
provides the time evolution, so that
\[ f(p,q,t) = e^{-i t \hat{L}} f(p,q,0) \] Note that any function
$g(H)$ is a constant of the motion, because $\hat{L} g(H) = 0$. The
time-evolution of macroscopic variables can then computed formally
from the microscopic processes captured by the Hamiltonian, by
considering the analytical properties of the Liouvillian. In
particular, we can express the mode auto-correlation explicitly in
terms of $\hat{L}$:
\begin{equation}
\chi_k (t) = \frac{\langle \delta n_k(0)\ e^{-i t \hat{L}} \delta n_k(0)
  \rangle}{\langle \delta n_k(0)^2 \rangle} 
\end{equation}

Extracting the mode-lifetime then becomes the challenge of
understanding how the Liouvillian couples the phase-space function
$\delta n_k$ to other modes. One approach constructed by Prigogine and
co-workers was to form a Taylor's series expansion in $t$, which can
be captured in a corresponding set of Feynman diagrams. Other
diagrammatic approaches have also been considered. These approaches
must all come to grips with the basic structure of the
auto-correlation. In the Fig.~\ref{fig:ChiAtOmega0}, it is clear that
the auto-correlation (once properly smoothed by a convenient choice of
transformation) is somewhat exponential in appearance ($\sim
e^{-|t|/\tau}$). However, $\chi(t)$ \emph{cannot} be simply
exponential. The requirement of time-reversal invariance would require
that the auto-correlation have zero slope at $t=0$ (see
Eq.~\ref{eq:chidot0}), but this is not consistent with purely
exponential behavior. Furthermore, a log-linear plot shows
clearly\cite{daw09b} that the behavior is not simply exponential, even
at moderate times. It might be argued that the very long-time behavior
should be exponential, but the mode lifetime (the area under the
auto-correlation) is not strongly influenced by the very long-time
tail of the auto-correlation, but by rather the moderate-time
behavior, where the behavior is not simply exponential. These
challenges have meant that the diagrammatic formalism has not been
very useful as a computational tool, in the sense that (at least no
our knowledge) no mode lifetimes for real systems have been calculated
using this formalism. The diagrammatic expansions are instructive for
general physical reasons, but we propose a different approach here,
which we believe will yield a more directly computable form. \\

Another approach would be to consider the eigenfunctions, or some
subset of the eigenfunctions, of the Liouvillian. Because $\hat{L}$ is
a linear (and also hermitian) operator, the properties of hermitian
operators and their eigenfunctions can be applied to the present case.
Related to that is the spectrum of $\hat{L}$. Generally speaking, that
is a tall task, but it is helpful to recognize that we are considering
only that part of phase space which is connected to $\delta n_k$ by
$\hat{L}$, and that simplifies the task. An efficient means for doing
this is based on the recursion method, which captures the structure of
the dynamics in an optimal way. We are therefore combining insights
from the recursion method with those from the work on the Liouvillian
in order to tackle the problem of the mode lifetime. \\

\section{The Recursion Method Applied to Mode Lifetimes}

Consider the auto-correlation of fluctuations in the auto-correlation
(see Eq.~\ref{eq:chik}), the area under which is the corresponding mode
lifetime $\tau$ (see Eq.~\ref{eq:GKform}). The general approach here
is to analyze $\tau$ in terms of the recursion scheme, as discussed at
length in the work of Haydock and
co-workers\cite{haydock80:_recur_method,haydock95,haydock99,haydock06}.
In principle, the recursion scheme provides a relevant analysis of the dynamics of 
$\delta n_k$. In that sense, the method identifies the important
sub-space of phase space, thereby focusing in on the relevant aspect
to be dealt with. The fluctuation $\delta n_k$ is a function on phase
space, and so is its time derivative
\[ \delta \dot{n}_k = i \hat{L} \delta n_k \] and successive
derivatives. The evolution in time can then be seen as a coupling via
the operation of $\hat{L}$ to other functions in phase-space. This
succession of states formally forms a linear
sequence the nature of which has been well-studied. \\

Consider then the sequence of functions in phase space generated by
acting on $\delta n_k$ with higher powers of $\hat{L}$. The first in
the sequence is $\delta n_k$ itself:
\[ v_0 = \delta n_k \]
and the next should reasonably be in the direction of $\delta \dot{n}_k$:
\[ v_1 = \hat{L} v_0 = -i \delta \dot{n}_k \] Now we need a measure
to determine to what extent the space has actually been expanded. This is
naturally provided by the ensemble average (see
Eq.~\ref{eq:ensembleavg}). Define the ``dot product'' of two phase
space functions as
\begin{equation} (A,B) = Q^{-1} \int d\Gamma\ e^{-\beta H}\ A^*\ B 
\label{eq:dotproduct}
\end{equation}
Then we can see that $v_1$ is orthogonal to $v_0$:
\[ (v_0,v_1) = (v_0, \hat{L} v_0) = (1, v_0 \hat{L} v_0) = \frac{1}{2}
(1, \hat{L} v_0^2) = \frac{1}{2}( \hat{L} 1, v_0^2) = 0 \] where the
second step is justified because $v_0$ is real, the fourth by the
hermiticity of $\hat{L}$, and the last because $\hat{L}$ acting on any
constant gives 0. (More generally, this can be seen as a consequence
of the time-reversal symmetry, Eqs.~\ref{eq:chidot0} and
~\ref{eq:nndot}). Normalizing, we have
\[ u_0 = \frac{\delta n_k}{\sqrt{ ( \delta n_k, \delta n_k )}} \]
\[ u_1 = \frac{ \hat{L} \delta n_k  }
{ \sqrt{ ( \hat{L} \delta n_k, \hat{L} \delta n_k ) } } \]
We note that $u_0$ is a real function and $u_1$ is pure imaginary.  \\

The sequence of orthonormalized functions is continued by 
the three-term recursion (taking $u_{-1} = 0$). 
\begin{equation}
b_{n+1} u_{n+1} = (a_n-\hat{L})u_n + b_n u_{n-1}
\label{eq:recursion}
\end{equation}
(along with the requirement that the $u_n$ are orthonormal). In the
first instance
\begin{equation}
b_1 u_1 = (a_0-\hat{L}) u_0
\label{eq:firstrecursion}
\end{equation}
along with $(u_0,u_1)) = 0$ gives $a_1 = (u_0, \hat{L} u_0) = 0$ where
the last result is from above. Taking the dot product of
Eq.~\ref{eq:firstrecursion} with itself and requiring $(u_1,u_1) = 1$
gives 
\[
b_1^2 = (\hat{L} u_0, \hat{L} u_0) = (u_0, \hat{L}^2 u_0)
\]

The next instance gives,
\[
b_2 u_2 = (a_1 - \hat{L}) u_1 + b_1 u_0
\]
We see that the choice above for $b_1$ already ensures that $(u_0,u_2)
= 0$. Requiring $(u_1,u_2) = 0$ gives $a_2 = (u_1, \hat{L} u_1)$.
We can repeat the argument that we used to show $a_1=0$ to show also
that $a_2=0$, except now we must remember that $u_1$ is pure
imaginary. Finally, requiring $(u_2,u_2) = 1$ gives
\[ b_2^2 = b_1^2 ( \frac{(u_0, \hat{L}^4 u_0)}{b_1^4} - 1 ) \] 
Finally we see that $u_2$ is pure real. \\

Continuing the sequence, we see that the functions $u_n$ are
alternately pure real and pure imaginary, that
\[ a_n = (u_n, \hat{L} u_n) = 0 \]
and that $b_n$ involves moments up to $(u_0, \hat{L}^{2 n} u_0)$. \\

The strength of this formalism is that this sequence of orthonormalized
functions is optimally configured to represent the dynamics of
$\delta n_k(t)$. Furthermore, in this
basis the representation of the Liouvillian is a special tridiagonal
form:
\[ \underline{\underline{L}} = \left[ 
\begin{array}{cccccccc}
0   & b_1 & 0   & \ldots & & & & 0 \\
b_1 & 0   & b_2 & 0 & \ldots & & & \\
0 & b_2 & 0 & b_3 & 0 & \ldots & & \\
. & . & . & . & . & . & . & . \\
. & . & . & . & . & . & . & .  \\
0 & . & . & . & . & . & . & . 
\end{array}
\right] \] so that the coefficients $b_n$ themselves provide the
representation. \\

The response of the system is extracted most easily in terms of the
resolvent of the Liouvillian
\[ \hat{R}(\omega) = (\omega - \hat{L})^{-1} \]
defined on the frequency domain.  The relevant behavior of the
resolvent is captured by the coefficients 
$b_n$. That is, we examine the projection of $R$ onto the initial
basis function:
\[ R_0(\omega) = (u_0,\hat{R}(\omega)u_0) = (\delta
n_k,(\omega-\hat{L})^{-1} \delta n_k)/(\delta n_k,\delta n_k) \] 
which can be expressed as the continued fraction 
\begin{equation}
R_0(\omega) = 1/\{ \omega - b_1^2/[\omega- \cdots -
b_n^2/(\omega- \cdots )]\} 
\label{eq:R0}
\end{equation}

\section{Partial Density of States}

The Partial Density of states (PDoS) $g_0(\omega)$ is given by the
imaginary part of the resolvent\cite{haydock80:_recur_method}
\[ g_0(\omega) = \frac{1}{\pi} | \Im\{R_0(\omega)\} | \]
and is related to the Fourier Transform (or power spectrum) $\tilde{\chi}_k(\omega)$ of 
$\chi_k(t)$ by
\[\tilde{\chi}(\omega) = \sqrt{2 \pi} g_0(\omega)\]
The PDoS is even and normalized and its moments 
are related to the coefficients $\{ b_k \}$ of the recursion, by
\[ \mu_m = \int_{-\infty}^{+\infty} d\omega\ \omega^m g_0(\omega) =
(u_0, \hat{L}^m u_0) \]
so that
\[ b_1^2 = \mu_2 \]
is related to the width of the distribution and the ratio
\[ \left( \frac{b_2}{b_1} \right)^2 = \frac{\mu_4}{\mu_2^2} - 1 \]
It is convenient to define 
\[ \gamma_4 \equiv \frac{\mu_4}{\mu_2^2} \] which is a low-order measure
of the shape of the power spectrum. A gaussian distribution has
$\gamma_4 = 3$, for example. In terms of $\gamma_4$, the angle $\theta$
between $\delta \ddot{n}$ and $\delta n$ is given by
\[ \cos{\theta} = \langle \delta \ddot{n} \delta n \rangle / \sqrt{
  \langle
  (\delta \ddot{n})^2 \rangle
 \langle \delta n^2 \rangle }  = -\gamma_4^{-1/2}
\] 

To summarize, the recursion method provides an optimally efficient
means of capturing all of the dynamical information relevant to the
dissipation of fluctuations in the aspect under consideration (in this
case, the occupation of a mode). The recursion maps the dynamics onto
a sequence of states which are coupled by the Liouvillian to the first
in the sequence, which is the occupation variable itself. The coupling
is evaluated in terms of moments of the Liouvillian, from which one
can reconstruct the resolvent and the auto-correlation function
itself. The net result is then that the mode lifetime, $\tau_k$ is a
function of the complete set of moments:
\begin{equation}
\tau_k = F(\mu_2, \mu_4, \mu_6, \ldots)
\label{eq:taufrommu}
\end{equation}
The function $F$ can be determined, in principle, from the resolvent.
However, finding the complete (infinite) set of
moments is impractical. In the next section, we discuss strategies for
dealing with knowledge of only a finite set. \\

\section{Reconstruction, Termination, and Convergence}

The problem then comes down to evaluating $\tau_k$ (or $g_0(\omega=0)$)
where we know some number of moments of the function
$g_0(\omega)$.  Based on dimensional analysis\cite{barenblatt96}, and noting
that the moment $\mu_m$ has dimension $\omega^m$, we can
re-write $\tau_k$ (Eq.~\ref{eq:taufrommu}) as a function
of all the moments as
\[
\tau_k = \mu_2^{-1/2} G( \gamma_4, \gamma_6, \ldots)
\]
so that the scale of $\tau_k$ is set by the second moment of the
distribution. The function G depends then on the dimensionless
parameters $\gamma_{2m} \equiv \mu_{2 m}/\mu_2^m$ which measure only
the relative \emph{shape} of the distribution. It is very unlikely
that we will ever find $G$ exactly, but using physical arguments we
should be able to find some close approximations to it. The hypothesis
proposed here is that a useful approximation can be obtained by
considering only a few low moments, the trade-off being that higher
moments are increasingly difficult to calculate. Some possible
approximations involving only low moments are discussed here.
Ultimately, the quality of these approximations will need to be judged
numerically (see\cite{daw09b}). \\

A simple approach would be to express the problem as one of
reconstructing the function from its low order moments, as one would
do in applying the principle of maximum entropy. In this fashion, we
determine $g_0(\omega)$ from its lowest non-vanishing $l$
moments by minimizing the functional \[ A\left[ g_0(\omega)
\right] = \int_{\infty}^{+\infty} d\omega\ \{g_0(\omega)\
ln{(g_0(\omega))} - \sum_{k=0}^{l-1}\ \lambda_k\ \omega^{2 k}
g_0(\omega)\} \] with respect to $g_0(\omega)$. At
the lowest (second moment) level, then, 
\[g^{(2)}_0(\omega) = \frac{1}{\sqrt{2 \pi} b_1} \exp{(\frac{-\omega^2}{2
    b_1})} \] and 
\[\tau_2^{ME} = \sqrt{2 \pi/\mu_2} \]
Incorporating higher moments involves exponentials of quartic and
higher powers in $\omega$. At the level of fourth moment, the spectrum will have the form
\[g^{(4)}_0(\omega) = A \exp{( - B \omega^2 - C \omega^4)} \] but the
relationship between the constants and the moments of the spectrum is
now transcendental, so that we cannot write an analytical form for
$B(b_1,b_2)$, for example. In that case, the relationship
between the lifetime and the moments is purely numerical. \\

It is helpful to note that we can incorporate more knowledge about the
resolvent than simply its lowest moments. We know that $\chi_k(t)$ is
infinitely differentiable, which means that $R_0(\omega)$ must fall
off faster than any power of $\omega$. Additionally, as pointed out by
Haydock, Nex, and Simons\cite{haydock99}, we know generally about the
analytical behavior of the projected $R_0(\omega)$ as a function of
complex $\omega$. Haydock and Nex\cite{haydock06} build on this
insight by proposing a general scheme for reconstructing the density
of states from the moments. They apply the physically motivated
constraint that states of the macroscopic system have minimal
lifetimes consistent with the moments, expressed alternatively as
maximal breaking of time-reversal symmetry (MBTS) in finite
subsystems. This they show can be expressed in terms of the tails
$T_N(\omega)$ of the continued fraction, defined at each level of
recursion by
\[
R_0(\omega) = 1/\{ \omega - b_1^2/[\omega- \cdots -
b_N T_N(\omega) ]\} 
\]
Because of the analytical properties of $T_N(\omega)$, inferred from
that of $R_0$, it is demonstrated that the continued fraction
expansion of $R_0$ converges within a circle of radius $\rho$ which
decreases exponentially with increasing $N$, which also produces the
same bounds on $g_0(\omega)$. The application of MBTS then constrains
the analytic properties sufficiently that a convergent calculation of
$g_0(\omega)$ is demonstrated. \\

At the lowest (second moment) level the power spectrum is Lorentzian, that is
\[ g^{(2)}_0(\omega) = \frac{b_1}{ \pi (\omega^2 + b_1^2)}\]
which gives 
\[ \tau_2^{MBTS} = 2/\sqrt{\mu_2} \]
However, the resulting auto-correlation function would be
$ \chi(t) = \exp{(-b_1 |t|)} $ which is not differentiable at $t=0$.
\\

The next (fourth moment level) gives
\[ g^{(4)}_0(\omega) = \frac{b_1 b_2 \sqrt{\omega^2 + b_1^2 + b_2^2}}
{ \pi [ (\omega^2-b_1^2)^2 + (\omega^2 + b_1^2) b_2^2 ] } \]
which gives
\[ \tau_4^{MBTS} = \tau_2^{MBTS} \sqrt{1-\gamma_4^{-1}}
\] 

Another approximation offered by Haydock and Nex~\cite{haydock06} is
the the single-band, where the spectrum $g_0(\omega)$ is non-zero only
inside of a finite range of frequency. In this case, the lowest level
gives
\[ g^{(2)}_0(\omega) = \frac{ \sqrt{4 b_1^2-\omega^2}}{2 \pi b_1^2} \]
for $| \omega | \leq 2 b_1$. This gives for the lifetime at the lowest
level of approximation  
\[ \tau_2^{SB} = 2/\sqrt{\mu_2} \]
The next level gives
\[ g^{(4)}_0(\omega) = \frac{ b_1 \sqrt{4 b_1 b_2^2-\omega^2}}
{2 \pi [ b_1^3 + (b_2-b_1) \omega^2 ] } \]
for $| \omega | \leq 2 \sqrt{b_1 b_2} $. Now the lifetime at the
second lowest level of approximation is
\[\tau_4^{SB} = \tau_2^{SB} \sqrt{\gamma_4-1} \]
We note that the single-band approximation will always give an
oscillatory $\chi(t)$. \\

Among these offerings, none is optimal. The maximum entropy approach
becomes analytically unwieldy at fourth moment. The MBTS leads to
power spectra which are power-law in $\omega$ and so do not fall off
fast enough, resulting in a $\chi(t)$ which has unphysical derivatives
at $t=0$. By contrast the SB gives oscillatory behavior for $\chi(t)$,
which is not appropriate for the examples above. However, they all
give somewhat similar dependence of $\tau$ on the lowest moments, and
it would seem that the final result is somewhat insensitive to the
details of the model. With some work on appropriate numerical
solutions, it may be possible to obtain a model for the power spectrum
which has all of the right properties and gives a usable but robust
expression for $\chi(t)$. \\

\section{Conclusions}

A formalism is presented here for improved calculations of vibrational
mode lifetimes in solids. The mode lifetime is calculated in terms of
moments of the power spectrum of the resolvent of the Liouvillian. The
lowest level (2nd moment) should provide a reasonably good
approximation to the lifetime, with refinement given by the next order
(4th moment). In the companion piece~\cite{daw09b}, we will compare
the approximations to the numerically exact calculation using the
Green-Kubo relation for some simple models of lattice vibrations. \\

\section{Acknowledgements}

This work was partly supported by DOE (\#DE-FG02-04ER-46139) and South
Carolina EPSCoR.

\newpage
%\bibliographystyle{unsrt}
%\bibliography{ILTC09_v2}

\end{document}